\documentstyle[twoside,fleqn,espcrc2,psfig]{article}

\newcommand{\AmS}{{\protect\the\textfont2

  A\kern-.1667em\lower.5ex\hbox{M}\kern-.125
  emS}}

\newcommand{\pabar}{\not{\!\partial}}

\newcommand{\Dbar}{\not{\!{\!D}}}

\title{Gravitino production during preheating and the equivalence theorem}

\author{A. L. Maroto and J.R. Pel\'aez\address{Dept. F\'{\i}sica Te\'orica,
Universidad Complutense de Madrid, 28040 Madrid, Spain }%
        \thanks{This work has been partially supported by
CICYT-AEN97-1693, PB-98-0782 and Ministerio de Educaci\'on y Ciencia (Spain).}}

\begin{document}

\begin{abstract}
We review our results on  the calculation of helicity $\pm 1/2$
gravitino production during preheating. The method we present is based on the
equivalence between goldstinos and longitudinal gravitinos at
high energies. The problem is thus reduced to the standard
(Majorana) fermion production after inflation. Comparison between
helicity $\pm 3/2$ and $\pm 1/2$ production and with
the results obtained in the unitary gauge is also presented.

\end{abstract}


\maketitle

\section{Introduction}
It is known that global supersymmetry allows the building of inflation models in
a relatively easy way. This is mainly due to the presence of scalars fields
in the theory and the fact that their potential terms typically contain
flat directions. Those flat directions, in addition, are not spoiled by radiative
corrections thanks to the non-renormalization theorems. However when dealing
with inflationary cosmology, global supersymmetry is not a good description and
gravitational effects should be introduced by promoting supersymmetry to
a local symmetry, i.e. supergravity. The supergravity corrections to the
inflaton potential can have very important effects, spoiling  its flatness
($\eta$ problem).
In addition, the appearance in the spectrum of very long-lived particles,
such as the gravitino, that couples only
with gravitational strength to the rest of matter fields, can pose new problems
during nucleosynthesis, since their decay products could destroy the nuclei
created in this period. This fact imposes a bound on the
abundance of gravitinos given by $n/s<10^{-14}$ for a gravitino
mass around $m_{3/2}\simeq 1$TeV. For this reason, it is very important
for the consistency of any model of inflation based on supergravity
to keep the gravitino production under control.

Gravitinos can be created by particle collisions in the thermal bath that appears after
inflation. The corresponding calculations of the particle production can be done by
standard perturbative techniques and leads to the well-known limit on the
reheating temperature $T_R<10^9$ GeV, for gravitinos with mass around 1 TeV.
However, gravitinos could also be created  directly from the
inflaton field during the period of its coherent
oscillations after inflation, provided the inflaton
field couples directly to the gravitino \cite{MaMa,Linde}. Under these conditions,
and during the first stages when
the amplitude of the oscillations is large, the perturbative technique is
no longer valid, and different non-perturbative approaches,
 such as Bogoliubov transformations
or lattice simulations, are required.

In this note we review our results on a new technique for the calculation
of the helicity $1/2$ gravitino production, based on the combination of
Bogoliubov transformations and the high-energy equivalence between
goldstinos and longitudinal gravitinos \cite{MaPe}.

\section{Gravitino helicities and the equivalence theorem}

Gravitinos are the spin $3/2$ supersymmetric partners of the graviton field.
As long as supersymmetry remains unbroken, the gravitino
is a massless field like the graviton  and, as a consequence, it
can only appear in two possible helicity states, namely, $\pm 3/2$.
However, when supersymmetry is spontaneously broken, the gravitino becomes a massive
field and then the $\pm1/2$ helicities are also possible. This  is
nothing but the supersymmetric version of the Higgs mechanism, although, instead of a
pseudoscalar massless
Goldstone boson "eaten" by a gauge boson that acquires a longitudinal component,
we now find a Majorana massless fermion, or goldstino, "eaten" by the gravitino
to acquire $\pm1/2$ helicity components. 
%This goldstino can be identified as the
%fermionic
%partnert of the scalar field that acquires the vacuum expectation value that breaks
%supersymmetry.

Intuitively, then, one is tempted to identify the  $\pm1/2$ helicity gravitino
components with those of the goldstino in a global supersymmetric theory. 
However, since those goldstinos are
massless and gravitinos are not, this intuitive identification can only be carried
out at energies high enough for the gravitino mass, $m$, to be negligible.
The rigorous formulation of this statement is known as the goldstino-gravitino
Equivalence Theorem. The use of this theorem
results in a major simplification, since it is
far simpler to work with Majorana fermions than with the Rarita-Schwinger spin 3/2 massive
gravitino projected along the $\pm1/2$ component.

Why $\pm1/2$ gravitinos are so relevant for preheating? The reason is
that the production of gravitinos with $\pm1/2$ helicities is, in general, dominant
during preheating.
This is due to the fact that we expect the physical momentum of the gravitinos,
$p_\mu$, to be of the order of the inflaton mass, $m_\phi$, which, typically,
is many orders of magnitude larger than the gravitino mass. If we now look at the
expression for the $\pm1/2$ helicity projectors, we find that they behave as
$P^{\pm1/2}_\mu\sim p_\mu/m+O(m/p)$, i.e., they grow with the momentum. In contrast, the
$\pm3/2$ projectors behave just like a constant of order one.
Consequently, if we want to calculate the number of gravitinos created after inflation,
we are basically dealing only with very energetic $\pm1/2$ gravitinos. Fortunately,
with the Equivalence Theorem are much easier to handle since they can be identified with
the Majorana
goldstinos.

\subsection{ Sketch of the proof}

The proof of the Equivalence Theorem, is somewhat technical, and we refer to our work for
mathematical details \cite{MaPe}. Our purpose here is to sketch the main ideas involved in its
derivation and
review the final result as well as 
the applicability conditions. Let us consider the fermionic part of the
minimal supergravity Lagrangian with a single chiral superfield which
contains an scalar field $\phi$ (inflaton) and
a Majorana spinor $\eta$ (inflatino, goldstino):
\begin{eqnarray}
g^{-1/2}{\cal L}_F&=&-\frac{1}{2} \epsilon^{\mu\nu\rho\sigma}
\bar\psi_\mu \gamma_5 \gamma_\nu D_\rho
\psi_\sigma\nonumber \\
&+&\frac{i}{2}\bar \eta
\Dbar \eta +e^{G/2}\left( \frac{i}{2}\bar \psi_\mu \sigma_{\mu\nu}
\psi^\nu\right. \nonumber \\&+&\left.\frac{1}{2}\left(-G_{,\phi\phi}-G_{,\phi}^2\right)
\bar \eta \eta \right. \label{lagrangiano} \\
&+&\left. \frac{i}{\sqrt{2}}G_{,\phi}\bar \psi_\mu
\gamma^\mu\eta\right)+\frac{1}{\sqrt{2}}\bar \psi_\mu
(\pabar \phi) \gamma^\mu
\eta\nonumber ,
\end{eqnarray}
where the K\"ahler potential is given by
$G(\Phi,\Phi^\dagger)=\Phi^\dagger\Phi+\log \vert W \vert^2$.
Note that the last two terms mix the gravitino,
$\psi_\mu$, and goldstino fields, and therefore their equations of motion
are coupled.

Let us then recall that supergravity is
a gauge invariant theory, invariant under local supersymmetry
transformations. Nevertheless, in order to make a specific calculation
it is necessary to choose a gauge, although the final result for the observables
will be independent of this choice. For instance, we could choose the unitary gauge,
where
we gauge away the goldstinos and make explicit the mass of the gravitino. In such a case
we would have to deal with the four helicities of the gravitino spin 3/2 field and 
therefore with the complicated $\pm1/2$ helicity projectors \cite{Linde}. 
But in any other gauge 
the goldstinos will still be
present in the Lagrangian, although there will be a mixing kinetic term between goldstinos
and the derivative of the gravitinos. This is the kind of gauges where the Equivalence
Theorem can be
derived \cite{Casalbuoni}.
Indeed, in these gauges, known as t'Hooft or $R_\xi$ gauges, the gauge fixing
function relates the goldstino to the {\it derivative} of the gravitino field over its
mass. The explicit gauge-fixing condition is given by:
\begin{eqnarray}
\gamma^\mu\psi_\mu
-\frac{1}{\sqrt{2}\xi\Dbar}e^{G/2}G_{,\phi}\eta
+\frac{ie^{-G/2}}{G_{,\phi}}\gamma^\mu(\pabar\phi)\psi_{\mu}=0\nonumber\\
\end{eqnarray}
that can be written asymptotically (before and after preheating) as:
\begin{eqnarray}
\partial^\mu\psi_\mu=\sqrt{\frac{3}{2}}\frac{m}{\xi}
\left(1-\xi\frac{m_{\pm}}
{m}\right)
\eta.
\label{gg}
\end{eqnarray}
Indeed, the derivative of the gravitino field is proportional
to the goldstino, with a proportionality constant that depends on
$\xi$.
Since, in momentum space, the derivative becomes the gravitino momentum, and we have
just seen that the  $\pm1/2$ helicity projector
$P_\mu^{\pm1/2}$ behaves precisely as the momentum over the mass, we find
that, in those gauges, the goldstinos are nothing but the $\pm1/2$ gravitinos, whenever
$p\gg m$. That is the standard proof of the Equivalence Theorem.

When dealing with the inflationary supergravity scenario, there are some subtleties
in this proof that we have addressed in our work \cite{MaPe}. 
Among others, it is specially relevant
the fact that we need to establish the equivalence in terms of classical equations of
motion in order to apply the Bogoliubov transformations.
However, previous proofs were given in terms of a path integral formulation or in
the context of
S-matrix theory \cite{Casalbuoni}. 
Further complications come from the definition of the asymptotic initial
and final states, the introduction of classical sources like curvature
and the inflaton itself, the fact that due to the expansion the initial and final
momenta are scaled with the Universe scale factor, etc... We refer to our work for the
details.

The important result is that in the asymptotic initial and final states, {\it the classical
solutions of the goldstino equations are proportional to  those of the $\pm 1/2$
helicity gravitinos}. Since the proportionality constant is {\it irrelevant} to
calculate the Bogoliubov coefficients (because they satisfy a normalization constraint),
the {\it production} of $\pm1/2$ gravitinos can be read directly from the calculation
using just goldstinos.

\subsection{Applicability conditions}

The previous statement can be applied only under the following conditions:
\begin{itemize}
\item The frequency of the inflaton oscillations  should be larger than the
gravitino mass $m_\phi\gg m$. If we are interested in pure gravitational production one
should also require $H\gg m$.
\item The classical sources should vanish asymptotically, which implies that both the
space time curvature as well as the inflaton oscillations should decrease with time.
\item  Since we are dealing with ensembles of gravitinos with a non-thermal distribution,
we have to be sure that the vast majority of the gravitinos are energetic enough, which
implies $a_{out} m \ll a_{in} m_\phi$, where $a_{in}$ and $a_{out}$ are the Universe
scale
factor before and after preheating, respectively.
\end{itemize}
The first, which we had already met, ensures that the typical gravitino momentum
is larger than its mass. The second is needed for the definition of particles in the
asymptotic regions. Concerning the third, the fact that there is an exponential growth
of the scale factor during inflation may seem to jeopardize the applicability of the
Theorem.
However, preheating only takes place during the first few inflaton oscillations, so that
the scale factor only grows by a few orders of magnitude, in contrast with the  many
orders of magnitude between $m$ and $m_\phi$, so that the Equivalence Theorem
 can be safely applied.

It is important to note that, {\it although the $\pm1/2$ helicity
gravitinos and the goldstinos may evolve differently} during the inflaton oscillations,
{\it they tend to each other in the asymptotic regions} (up to a proportionality constant).

\subsection{The calculation with goldstinos and the Landau gauge}

Once we have established that we simply need to study the gravitino production
in terms of the much simpler {\it goldstino fields},
next, we have to calculate the evolution of goldstino
plane waves under the {\it goldstino equations} of motion, and compare the particle number
in the initial and final states.

The problem is that even though we are simply looking at Majorana fermions,
their equations of motion are still rather complicated in a general $R_\xi$ gauge.
From eq.(\ref{lagrangiano}) we obtain:
\begin{eqnarray}
&&i\Dbar \eta -e^{G/2}\left(G_{,\phi\phi}+G_{,\phi}^2\right)\eta\nonumber \\
&&\hspace{1cm}-e^{G/2}G_{,\phi}\frac{i}{2\xi\Dbar}e^{G/2}G_{,\phi}\eta=0
\label{goldxi}
\end{eqnarray}
However, we still have the freedom to choose the gauge fixing variable $\xi$ to obtain
a further simplification. The most convenient choice seems to be $\xi\rightarrow\infty$,
known as the Landau gauge, since then the goldstinos obey a well known Dirac-like
equation, which can be easily integrated numerically.

\section{Numerical example with a toy model}

In order to illustrate the above analysis we will consider a particular
example with a superpotential given by:
\begin{eqnarray}
W=\sqrt{\lambda}\frac{\Phi^3}{3},
\label{super}
\end{eqnarray}
According to the Equivalence Theorem, the number of helicity $\pm 1/2$
gravitinos created by
the oscillating inflaton
can be obtained by solving the equation for the Fourier modes of the goldstinos
coming from (\ref{goldxi}). Using the standard Bogoliubov coefficients technique we
obtain the results for the spectrum shown in Fig.1. (The helicity $\pm 3/2$ results
have been obtained following the analysis in \cite{MaMa}).
\begin{figure}[htbp]
\vspace*{-.5cm}
\hspace*{-.5cm}
%\hbox{\psfig{file=comp001-1.ps,width=9cm}}
\hbox{\psfig{file=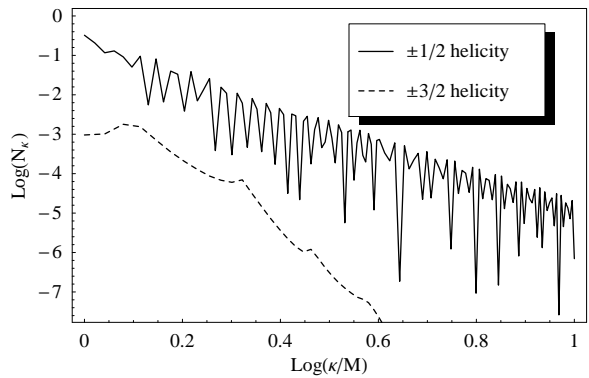,width=8cm}}

\vspace*{-1.cm}

\caption{Spectrum of  helicity $\pm 3/2$ and helicity $\pm 1/2$
gravitinos. The helicity $\pm 1/2$
production has been obtained using the
ET.}
\end{figure}
We see that for this particular model, the production of helicity
$\pm 3/2$ gravitinos is suppressed by two to three orders of magnitude
with respect to the $\pm 1/2$ gravitinos. Integrating the spectrum over the
whole range of momenta we obtain the total number of gravitinos
created during preheating. For the helicity $\pm 1/2$ we obtain
$n/s\simeq 10^{-10}$. Comparing this result with the nucleosynthesis bounds
we observe that this model would be ruled out if the gravitino mass is around (or smaller
than) 1 TeV.
We have found a very good agreement between 
these results and those obtained in the unitary gauge \cite{Linde} for the same model.

\section{Future and Possible extensions}

In the previous discussion we have concentrated on the simplest case in which we
have minimal supergravity and only one chiral superfield. We should recall that
apart from a successful inflationary period, we also require from our model that
supersymmetry is broken at the minimum of the potential, otherwise the production
of helicity $\pm 1/2$ gravitinos would be meaningless.
Precisely for this reason, the single-field theories cannot be considered, in general,
as realistic models  for supergravity inflation
since, on the one hand, they typically exhibit the $\eta$ problem described in the introduction,
and, on the other hand, it is extremely difficult to find an appropriate 
superpotential that breaks supersymmetry at the minimum and, at the same time,
contains the appropriate inflationary scale. It is then very interesting to
extend the analysis to the multi-chiral fields models in which it is possible to accommodate
these requirements \cite{Linde2}. In this case, although the identification
of the goldstino is not straightforward and in general it will
 be a time-dependent combination of fermionic fields, we  expect on physical grounds
an extended version of the Equivalence Theorem will also hold. Work is in progress in
this direction.

\end{document}